\documentclass[final,3p,times,twocolumn]{elsarticle}
\usepackage{graphicx}
\usepackage{amssymb} 
\usepackage{amsmath} 

\usepackage{lineno}

\usepackage{float} 
\usepackage{subfig} 

\usepackage{ifpdf}
\ifpdf
\usepackage[pdftex]{hyperref}
\else
\usepackage[hypertex]{hyperref}
\fi

\hypersetup{
  pdftitle={},%
  pdfauthor={},%
  pdfsubject={},%
  pdfkeywords={},%
  pdfstartview={},%
  bookmarksopen=true, breaklinks=true, debug=true, %
  colorlinks=true, linkcolor=red, citecolor=blue, urlcolor=blue
}

\journal{elsevier: SLAC-PUB-13530}

\newcommand{\beq}{\begin{eqnarray}}
\newcommand{\eeq}{\end{eqnarray}}
\newcommand{\be}{\begin{eqnarray*}}
\newcommand{\ee}{\end{eqnarray*}}
\newcommand{\nn}{\nonumber}

\newcommand{\cc}{{c\bar{c}}}

\newcommand{\vect}[1]{\vec{#1}}

\newcommand{\eqs}[1]{\begin{equation} \begin{split} #1\end{split} \end{equation} }

\newcommand{\ie}{{\it i.e.}}

\newcommand{\F}{\mathcal F}
\newcommand{\R}{\mathcal R}

\newcommand{\ce}[1]{Eq.~\eqref{#1}}

\newcommand{\cf}[1]{{Fig.~\ref{#1}}}

\def\lsim{\raise0.3ex\hbox{$<$\kern-0.75em\raise-1.1ex\hbox{$\sim$}}}
\def\gsim{\raise0.3ex\hbox{$>$\kern-0.75em\raise-1.1ex\hbox{$\sim$}}}

\def\CuCu {CuCu}

\def\dAu  {$d$Au}

\def\AuAu {AuAu}
\def\pp   {$pp$}
\def\pA   {$pA$}
\def\AA   {$AA$}
\def\AB   {$AB$}

\def\sqrtsNN {\mbox{$\sqrt{s_{NN}}$}}
\def\Npart   {\mbox{$N_{\rm part}$}}
\def\Ncoll   {\mbox{$N_{\rm coll}$}}

\def\RdAu    {\mbox{$R_{d\rm Au}$}}
\def\RAuAu    {\mbox{$R_{\rm AuAu}$}}
\def\RCuCu    {\mbox{$R_{\rm CuCu}$}}

\def\jpsi    {\mbox{$J/\psi$}}
\def\pT      {\mbox{$P_{T}$}}
\def\beq     {\begin{equation}}
\def\eeq     {\end{equation}}

\long\def\symbolfootnote[#1]#2{\begingroup%
  \def\thefootnote{\fnsymbol{footnote}}\footnote[#1]{#2}\endgroup}

\begin{document}


\begin{frontmatter}

\title{{Cold nuclear matter effects on $J/\psi$ production:\\
intrinsic and extrinsic transverse momentum effects}}

\author[santiago]{E. G. Ferreiro}
\author[LLR]{F. Fleuret}
\author[SLAC,heidelberg]{J.P. Lansberg\fnref{present}}
\author[CEA]{A. Rakotozafindrabe}

\address[santiago]{Departamento de F{\'\i}sica de Part{\'\i}culas, 
Universidad de Santiago de Compostela, 15782 Santiago de Compostela, Spain}
\address[LLR]{Laboratoire Leprince Ringuet, \'Ecole Polytechnique, CNRS-IN2P3,   91128 Palaiseau, France}
\address[SLAC]{SLAC National Accelerator Laboratory, Theoretical Physics, Stanford University, Menlo Park, CA 95025, USA 
}
\address[heidelberg]{ Institut f\"ur Theoretische Physik, Universit\"at Heidelberg, 
Philosophenweg 19,  D-69120 Heidelberg, Germany}
\address[CEA]{IRFU/SPhN, CEA Saclay,
  91191 Gif-sur-Yvette Cedex, France}

\fntext[present]{Present address at SLAC.}

\author{}

\address{}

\begin{abstract}
\small
Cold nuclear matter effects on \jpsi\ production in proton-nucleus and 
nucleus-nucleus collisions are evaluated taking into account the specific 
\jpsi-production kinematics at the partonic level, the shadowing 
of the initial parton distributions and the  absorption in the nuclear 
matter. We consider two different parton processes for the  $c \bar c$-pair 
production: one with collinear gluons and a recoiling gluon in the final 
state and the other with initial gluons carrying intrinsic transverse momentum.
Our results are compared to RHIC observables. The smaller values of the nuclear 
modification factor $R_{AA}$ in the forward rapidity region (with respect to the 
mid rapidity region) are partially explained, therefore potentially reducing the 
need for recombination effects.
\end{abstract}


\end{frontmatter}


\section{Introduction}
\label{sec:intro}

The charmonium production in hadron collisions is an on-going major subject of 
investigations, on both experimental and theoretical sides. It has been widely
studied in \pp\ collisions; our understanding was recently 
reviewed in~\cite{Lansberg:2006dh,Lansberg:2008gk}. It may also be used as a tool to probe 
the medium produced in nucleus-nucleus (\AB) collisions (for a recent 
review, see~\cite{Rapp:2008tf} along with some perspectives for the LHC~\cite{Lansberg:2008zm}). 
This medium is expected 
to be in a deconfined state of QCD matter -- such as the Quark Gluon Plasma (QGP) -- 
at high enough temperatures and densities. The \jpsi\ production should be sensitive to 
the QGP formation, due to compe\-ting effects such as a color Debye screening 
suppression~\cite{Matsui86} or the so-called recombination mechanism~\cite{recombniationRefs}. 
Recent results on  \jpsi\ production are available 
from the PHENIX experiment at the BNL Relativistic Heavy Ion Collider (RHIC). They show a significant 
suppression of the \jpsi\ yield in \AuAu\ collisions at 
$\sqrtsNN=200\mathrm{~GeV}$~\cite{Adare:2006ns} compared to the expected yield 
from \pp\ measurements~\cite{Adare:2006kf}. However, the interpretation relies on a good 
understan\-ding and a proper subtraction of the Cold Nuclear Matter (CNM) effects, known to 
impact the \jpsi\ production in proton(deuteron)-nucleus ($pA, dA$) collisions where the 
deconfinement can not be reached. Indeed, the $pA$ data~\cite{Alessandro:2006jt} obtained at 
the SPS energies can be described by assuming the break-up of the pre-resonant $\cc$~pair due 
to multiple scattering along its way to escape the nuclear environment -- the so-called nuclear 
absorption. PHENIX data on \dAu\ collisions~\cite{Adare:2007gn} have also revealed 
that CNM effects play an essential role at RHIC energy. These effects {\it a priori} include 
shadowing, \ie~the modification of the parton distribution of a nucleon in a nucleus,  
and final-state nuclear absorption.

As we shall show thereafter,  the impact of gluon shadowing depends on the partonic 
process producing the $c \bar c$ and then the \jpsi. So far, the 
studies of \jpsi\ production~\cite{OtherShadowingRefs,Vogt:2004dh,OurIntrinsicPaper} with
gluon shadowing relied on the assumption that the $c \bar c$  pair was produced by the fusion of two gluons
 carrying some intrinsic transverse momentum~$k_T$. The partonic process being a $2\to 1$ scattering, 
the sum of the gluon intrinsic transverse momentum is transferred to the $c \bar c$ pair, thus to 
the \jpsi\ since the soft hadronisation process does not modify the kinematics. This corresponds 
to the picture of the Colour Evaporation Model (CEM) at LO (see ~\cite{Lansberg:2006dh} and references 
therein).

In such approaches, the transverse momentum of the \jpsi\ {\it entirely} comes from the intrinsic 
transverse momentum of the initial gluons. This seems acceptable for the low-$P_T$ region and the
origin of this intrinsic $P_T$ can be paralleled to the increase of $\langle P_T^2\rangle$ when going from 
\pp\ to \pA\, and for increasing atomic number~$A$. This is known as the Cronin effect: the increase
 of $\langle P_T^2\rangle$ is believed to come from a broadening of the intrinsic transverse momentum 
distribution, resulting from the multiple scatterings experienced by the initial gluon from the proton as it 
goes through the target nucleus before 
the heavy-quark production~\cite{Hufner:1988wz}. 

However, such an effect is not sufficient to describe the $P_T$ spectrum of quarkonia produced in 
hadron collisions~\cite{Lansberg:2006dh}. Most of the transverse momentum should have an extrinsic 
origin, \ie\ the \jpsi's $P_T$ would be balanced by the emission of a recoiling particle in the final 
state. The \jpsi\ would then be produced by gluon fusion and with emission of a hard final-state gluon. 
For the production of ${}^3S_1$ states -- like the \jpsi, such a $2\to 2$ partonic process is 
anyhow mandatory to sa\-tisfy $C$-parity conservation. 

It is among our purposes here to investigate the influence of such an emission on the kinematics of the 
\jpsi\ production in $p(d)A$ and $AA$ collisions. Indeed, for a given \jpsi\ momentum (thus for 
fixed~$y$ and $P_T$), the processes discussed above, \ie\ $g+g \to c\bar c \to J/\psi \,(+X)$  
and $g+g \to J/\psi +g$,  will proceed on the average from gluons with different Bjorken-$x$. Therefore,
 they will be affected by different shadowing corrections. From now on, we will refer to the former 
scenario as the {\it intrinsic} scheme, and to the latter as the {\it extrinsic} scheme. In the 
following, we shall consider them as distinct approaches\footnote{In the extrinsic scheme and for this first study, we shall
neglect the small kinematical effects of the yield from the decay of $\chi_c$ produced by $2\to 1$ processes. Indeed,
from the 25~\% of the $\chi_c$ feeddown~\cite{Faccioli:2008ir}, it is reasonable to suppose
that only half of it effectively proceeds via a $gg\to\chi_c$ which is allowed at LO in $\alpha_S$
only for $\chi_{c2}$, not for $\chi_{c1}$.}.

In practice, we shall study the kinematic regime at RHIC at $\sqrtsNN=200\mathrm{~GeV}$. 
For the extrinsic scheme, we shall consider the partonic differential cross section for 
$g+g\to J/\psi+ g$ given in~\cite{Haberzettl:2007kj} which satisfactorily describes the data 
obtained in \pp\ collisions at RHIC down to $P_T\sim 0$ (see \cf{fig:extrinsicpT}). For the intrinsic scheme, we shall 
follow the studies~\cite{OtherShadowingRefs,Vogt:2004dh,OurIntrinsicPaper} based on ($2\to 1$)-like 
processes where the momentum of the particles denoted by $X$ in
$g+g \to c\bar c \to J/\psi \,(+X)$ is neglected . Using a 
probabilistic Glauber Monte-Carlo code, we interface these production processes with CNM effects, such 
as shadowing and nuclear absorption, in order to get the \jpsi\ production cross sections 
for \pA\ and \AA\ collisions. We shall finally compare our results with the experimental 
measurements presently available at RHIC.

\begin{figure}[thb!]
\begin{center}
\subfloat[][$P_T$ distributions from the $s$-channel cut contributions compared to PHENIX \pp\ data~\cite{Adare:2006kf} in the forward and central rapidity regions.]{
\label{fig:extrinsicpT-a}
\includegraphics[height=6cm,angle=90]{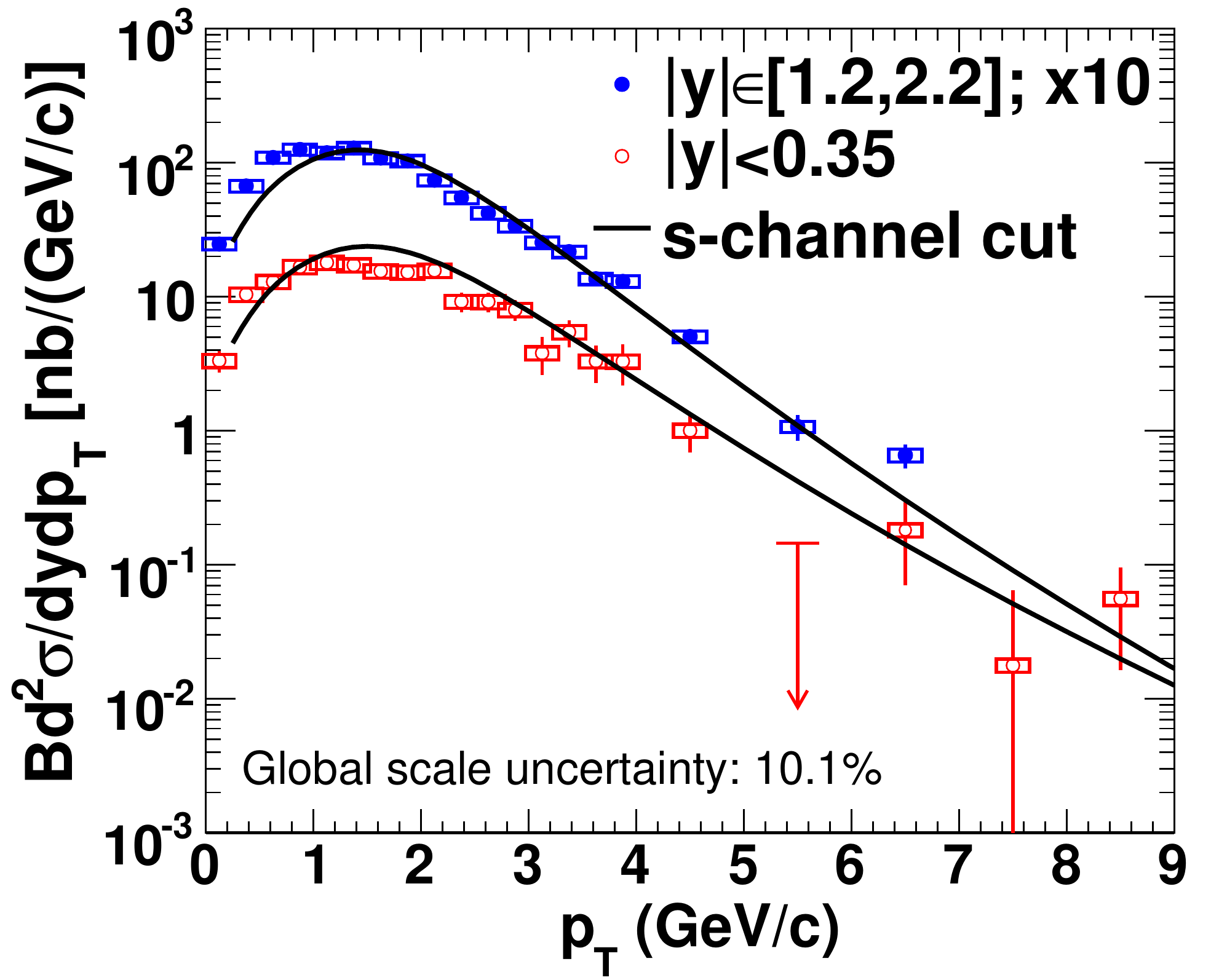}}
\\
\subfloat[][Rapidity spectrum from $s$-channel cut 
contributions compared to PHENIX \pp\ data, the {\it ad hoc} double-Gaussian fit~\cite{Adare:2006kf}
and the predictions from the PYTHIA event generator.]{
\label{fig:extrinsicpT-b}
\includegraphics[width=4.3cm,height=7.3cm,angle=90]{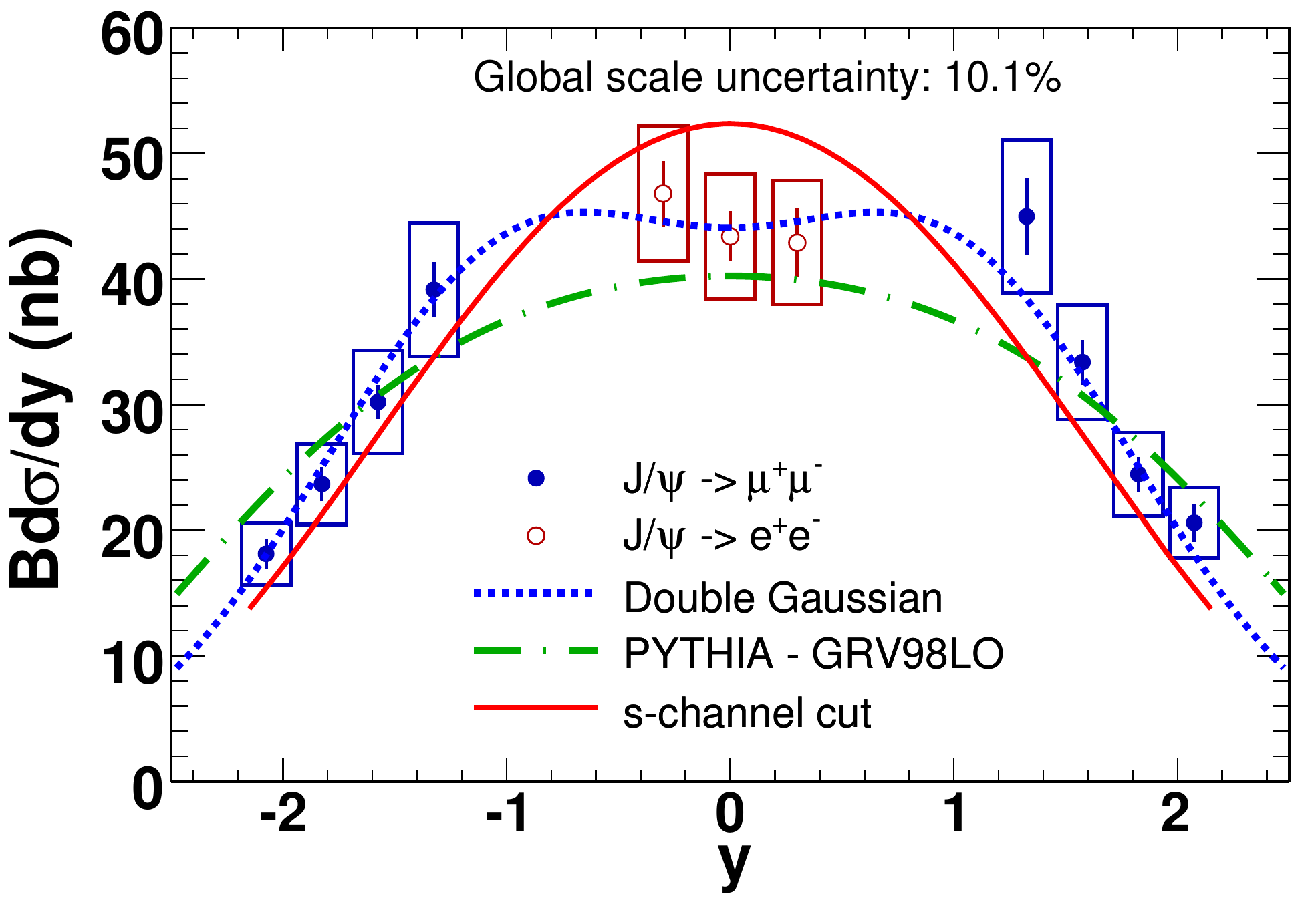}}
\end{center}
\caption{$P_T$ and $y$ spectra in \pp\ at $\sqrtsNN=200\mathrm{~GeV}$.}
\label{fig:extrinsicpT}
\end{figure}


\section{The Monte-Carlo framework for \jpsi\ production}
\label{sec:ourapproach}


To describe the \jpsi\ production in nucleus collisions, our Monte-Carlo 
framework is based on the probabilistic Glauber model, the nuclear density
 profiles being defined with the Woods-Saxon parameterisation for
any nucleus ${A>2}$ and the Hulthen wavefunction for the 
deuteron~\cite{Hodgson:1971}. The nucleon-nucleon inelastic cross section at
$\sqrtsNN=200\mathrm{~GeV}$ is taken to $\sigma_{NN}=42\mathrm{~mb}$ and the 
maximal nucleon density to $\rho_0=0.17\mathrm{~nucleons/fm}^3$. For each event 
(for each \AB~collision) at a random impact parameter~$b$, the Glauber Monte-Carlo 
model allows us to determine the number of nucleons in the path of each incoming 
nucleon, therefore allowing us to easily derive the number \Ncoll\ of 
nucleon-nucleon collisions and the total number \Npart\ of nucleons 
participating into the collision.

\subsection{The CNM effects}
\label{subsec:th-CNM-effects}

To get the \jpsi\ yield in \pA\ and \AA\ collisions, a shadowing-correction 
factor has to be applied to the \jpsi\ yield obtained from the simple 
superposition of the equivalent number of \pp\ collisions.
This shadowing factor can be expressed in terms of the ratios $R_i^A$ of the 
nuclear Parton Distribution Functions (nPDF) in a nucleon of a nucleus~$A$ to the 
PDF in the free nucleon. In the following, we will consider the evolution model 
EKS98~\cite{Eskola:1998df} which is recognised to be a reasonable compromise
between DS~\cite{deFlorian:2003qf} and EPS08~\cite{Eskola:2008ca} as regards 
the strength of the gluon antishadowing for instance.  
It provides $R_i^A$ at a given initial value of $\mu_F$ -- the factorisation 
scale -- and takes into account their evolution through the DGLAP
equations. The nuclear ratios of the PDFs are expressed by:
\beq
\label{eq4}
R^A_i (x,\mu_F) = \frac{f^A_i (x,\mu_F)}{ A f^{nucleon}_i (x,\mu_F)}\ , \ \
f_i = q, \bar{q}, g \ .
\eeq
Within EKS98, these nuclear ratios are parameterized at some initial scale 
$\mu^2_{F,0} = 2.25\mathrm{~GeV^2}$ which is assumed large enough for perturbative 
DGLAP evolution to be applied. They are evolved at LO from $\mu^2_{F,0}$ up to $\mu_F^2$ 
($<10^4\mathrm{~GeV^2}$) 
and are valid for $x \geq 10^{-6}$. The nu\-merical parameterisation of $R_i^A(x,\mu_F)$ 
is given for all parton flavours. Here, we restrain our study to gluons since, at 
high energy, \jpsi\ is essentially produced through gluon fusion \cite{Lansberg:2006dh}. 
Usually, the spatial dependence of $R_i^A(x,\mu_F)$ is not given. However, as we shall 
see in the next section, it can be included in our approach.

The second CNM effect that we are going to take into account concerns 
the nuclear absorption.  In the framework of the probabilistic Glauber 
model, this effect refers to the probability for the pre-resonant $c{\bar c}$ 
pair to survive to the propagation through the nuclear medium and is usually parametrised
by introducing an effective absorption cross section~$\sigma_{\mathrm{abs}}$.

\subsection{The differential \jpsi-production cross section in \AB~collisions}
\label{subsec:diff-xsection}

Our Glauber Monte-Carlo framework is aimed at numerically evaluating the 
differential \jpsi-production cross section in nucleus collisions, by 
exploring the whole physical phase space with a proper weighting of each 
point in this phase space. Within this Glauber-based calculation, the 
differential cross section for the production (via gluon fusion) of a 
quarkonium with momentum $(y,P_T)$, in nucleus collisions at impact 
parameter $\vect b$ and for a  nucleon-nucleon CM energy of \sqrtsNN\ 
can be represented by a generic integral. It takes two different forms 
depending on the kinematics of the partonic process responsible for the 
\jpsi\ production. 

The intrinsic scheme corresponds to a $2\to 1$ partonic process, with 
initial gluons carrying a non-zero intrinsic transverse momentum. 
Following~\cite{OurIntrinsicPaper}, we do not neglect the value of the
 \jpsi's $P_T$ in this simplified kinematics. In this scheme, the 
measurement of the \jpsi\ momentum completely fixes the longitudinal
 momentum fraction carried by the initial partons:
\begin{equation}
x_{1,2} = \frac{m_T}{\sqrt{s_{NN}}} \exp{(\pm y)} \equiv x_{1,2}^0(y,P_T),
\label{eq:intr-x1-x2-expr}
\end{equation}
with the transverse mass $m_T=\sqrt{M^2+P_T^2}$, $M$ being the \jpsi\ mass. 

Therefore, we can write
\begin{align}
\label{eq:intr_master}
&\frac{d \sigma^{\rm Intr.}_{AB}}{dy \, dP_T \, d\vect b } =  \int d\vect r_A \, dz_A \, dz_B \, \F^A_g(x^0_1,\vect r_A, z_A,\mu_F)\ 
\\&\F^B_g(x^0_2,\vect r_B, z_B,\mu_F)
\,\sigma^{\rm Intr.}_{gg}(x^0_1,x^0_2) \,S_A(\vect r_A,z_A) \,S_B(\vect r_B,z_B)\nn \, 
\end{align}
with
\begin{itemize}
\item 
$\vect r_A$ [$\vect r_B=\vect b-\vect r_A$] and $z_A$ [$z_B$] are the transverse and longitudinal spatial locations of the initial parton (a gluon here) in the nucleus~$A$~[$B$]; it carries the longitudinal momentum fraction $x^0_1$ [$x^0_2$] and can be found with the probability $\F^A_g$ [$\F^B_g$] at a scale $\mu_F$;
\item The nuclear absorption is taken into account through 
$S_A(\vect r_A, \,z_A)= \exp \left ( - A \,\sigma_{\mathrm{abs}}
\int_{z_A}^{\infty} d\tilde{z}\ \rho_A(\vect r_A, \tilde{z}) \right )$ [$S_B(\vect r_B, \,z_B)$], which stands for the survival probability for a $c\bar{c}$
produced at the point $(\vect r_A,z_A)$ [$(\vect r_B,z_B)$] to pass through the projectile and 
the target unscathed;
\item $\sigma^{\rm Intr.}_{gg}(x^0_1,x^0_2)$ is the partonic cross section
for the process $g+g\to c\bar c \to J/\psi(+X)$ and is a function of $P_T$ and $y$ through $x^0_1$ and $x^0_2$. As we will show in the section~\ref{subsec:partonic-xsection}, it can
be extracted from experimental data along with the PDFs.
\end{itemize}

In the extrinsic scheme, we deal with a  $2\to 2$ partonic process with collinear initial gluons and we have
\begin{align}
&\frac{d \sigma_{AB\to J/\psi X}}{dy \, dP_T \, d\vect b }=
\!\int \!\!dx_1 dx_2 \int d\vect r_A dz_A  dz_B   \F^A_g(x_1,\vect r_A, z_A,\mu_F)\nn\\
&\F^B_g(x_2,\vect r_B, z_B,\mu_F)
2 \hat s P_T\frac{d \sigma_{gg\to J/\psi + g}}{d \hat t} 
\delta(\hat s-\hat t-\hat u-M^2)\nn
\\&S_A(\vect r,z_A) S_B(\vect r_B,z_B) \ ,
\end{align}
with
\begin{itemize}
\item $\hat s= s_{NN} x_1 x_2,\  \hat t= M^2-x_1 \sqrt{s_{NN}} m_T e^{y},\ \hat u= M^2-x_2 \sqrt{s_{NN}} m_T e^{-y}$. 
The four-momentum conservation -- represented by the $\delta$ function -- explicitly results in a more 
complex expression of $x_2$ as a function of~$(x_1,y,P_T)$:
\begin{equation}
x_2 = \frac{ x_1 m_T \sqrt{s_{NN}} e^{-y}-M^2 }
{ \sqrt{s_{NN}} ( \sqrt{s_{NN}}x_1 - m_T e^{y})} \ .
\label{eq:x2-extrinsic}
\end{equation}  
Equivalently, a similar expression can be written for $x_1$ as a function of~$(x_2,y,P_T)$;
\item $d \sigma_{gg\to J/\psi +g} / d \hat t$ can be computed in {\it a priori} different approaches that 
correctly describes the \pp\ data. For now, we use the one obtained in~\cite{Haberzettl:2007kj}, but others can be interfaced with our code. See section~\ref{subsec:partonic-xsection} for further details.
\end{itemize}

Now, concerning $\F^A_g(x_1,\vect r_A, z_A,\mu_F)$, 
we assume that it can be factorised in 
the nuclear density distribution $\rho_A(\vect r_A,z_A)$, the shadowing modification factor 
$\R^A_g (\vect r_A,x,\mu_F)$ and the usual gluon PDFs $g(x;\mu_F)$:
\begin{equation}
\F^A_g(x_1,\vect r_A, z_A;\mu_F) = \rho_A(\vect r_A,z_A)  \: \R_g^A(\vect r_A,x_1,\mu_F) \: g(x_1;\mu_F) \ .\nn
\end{equation}

{\it A priori}, the modifications of the nPDFs should depend on the parton position~$(\vect r, z)$ in the nucleus. 
Such information is not experimentally available. So the centrality dependence is not encoded in the EKS98~parametrisation. However, some approaches provide an Ansatz for such a dependence.
Assuming that the inhomogeneous shadowing is proportional to the path length~\cite{Klein:2003dj,Vogt:2004dh}, then

\beq
\label{eq:inhomogeous-shadow}
\R^A_g (\vect r_A,x,\mu_F)=1+[R^A_g (x,\mu_F)-1]N_{\rho_A}
\frac{\int dz \,\rho_A(\vect r_A,z)}{\int dz \,\rho_A(0,z)}
\eeq
where $R^A_g (x,\mu_F)$ is the ratio nPDF/PDF given by the EKS98 parametrisation for the gluon (see Eq.~\eqref{eq4}) and $N_{\rho_A}$ 
is a normalisation factor,  determined such that 
\eqs{
\frac{1}{A} \int d^2\vect r_A \int dz_A \rho_A(\vect r_A,z_A) \R^A_g (\vect r_A,x,\mu_F) = R^A_g (x,\mu_F) \ .\nn}

The integral over $z$ in Eq.~\eqref{eq:inhomogeous-shadow} includes all the material traversed by the incident nucleon. 
This amounts to consider the incident parton as coherently interacting with all the target partons along its path length.

In the following, we shall set the scale $\mu_F$ in $\F^A_g$ equal to the renormalisation scale $\mu_R$ of the partonic process and take the usual choice $\mu_R=m_T$, $m_T$ providing a ty\-pi\-cal scale of
the partonic process.

\subsection{The partonic cross sections}
\label{subsec:partonic-xsection}

Within the intrinsic scheme, the measurement of the differential cross section 
in \pp\ on \cf{fig:extrinsicpT} (a) and (b) directly provides us with values for the product of the PDFs and the 
partonic cross section in \ce{eq:intr_master}. Indeed,
\eqs{
\frac{d\sigma^{\rm Intr.}_{pp}}{dy \, dP_T} = g(x_1^0;m_T) \, g(x_2^0;m_T) 
\, \sigma^{\rm Intr.}_{gg}(x^0_1,x^0_2) \ .}

In order to evaluate the integral \ce{eq:intr_master}, we randomly pick $y$ and \pT\ out of their 
respective distributions obtained from fits to the experimental \pp~spectra $d\sigma^{\rm Intr.}_{pp}/dy \, dP_T$ . 
For the present study, we use the fits to the $y$~\footnote{The double gaussian parametrisation as quoted in~\cite{Adare:2006kf}.}
 and \pT\ spectra measured 
by PHENIX~\cite{Adare:2006kf} in \pp\ collisions at $\sqrt{s_{NN}}=200\mathrm{~GeV}$ 
as inputs of the Monte-Carlo. The azimuthal-angle $\varphi$ of \pT\ in the 
$(P_x,P_y)$ plane is also random and follows a flat distribution within 
$[0,2\pi]$. As discussed previously, the know\-ledge of $(y, P_T)$ unequivocally
 fixes the other va\-ria\-bles ($x_1$, $x_2$ and $\mu_F=m_T$) needed to compute 
the shadowing correction factors.

On the other hand, in the extrinsic scheme, information from the data alone 
-- the $y$ and \pT\ spectra -- is not sufficient to determine $x_1$ and $x_2$. 
Indeed, the presence of a final-state gluon authorises much more freedom to 
choose $(x_1, x_2)$ for a given set $(y, P_T)$. Even if kinematics determine 
the physical phase space, models are anyhow mandatory to compute the proper 
weighting of each kinematically allowed $(x_1, x_2)$. This weight is simply 
the differential cross section at the partonic level times the gluon PDFs, 
\ie\ $g(x_1,\mu_F) g(x_2, \mu_F) \, d\sigma_{gg\to J/\psi + g} /dy \, dP_T\, dx_1 dx_2 $. 
In the present implementation of our code, we are able to use the partonic differential 
cross section computed from {\it any} theoretical approach. For now, we use the one 
from~\cite{Haberzettl:2007kj} which takes into account the $s$-channel cut 
contributions~\cite{Lansberg:2005pc} to the basic Color Singlet Model (CSM)~\cite{CSM_hadron} and 
satisfactorily describes the data down to very low~\pT, where the bulk of 
the cross section lies. As shown on Fig.~\ref{fig:extrinsicpT}, this 
approach\footnote{The determination of the two parameters of this approach~\cite{Haberzettl:2007kj}
 has been improved by fitting RHIC \pp\ data ($a=3.2$ and $\kappa=6.3\mathrm{~GeV}$).} gives 
a fairly good description of both the \pT\ and $y$ dependence of the \pp\ data at RHIC.

To evaluate the integral in the extrinsic scheme, we also randomly pick $y$ and \pT, but 
out of the distributions computed with the cross section computed as in~\cite{Haberzettl:2007kj}. 
For a given set $(y, P_T)$, 
the set $(x_1,x_2)$ is randomly chosen in its kinematically allowed range and follows the 
distribution $g(x_1,\mu_F) \, g(x_2, \mu_F) \, d\sigma_{gg\to J/\psi + g} /dy \, dP_T dx_1 dx_2$.

\subsection{Picturing intrinsic vs extrinsic scheme}
\label{subsec:see-phase-space}

\cf{fig:intriextriX} shows the physical phase space in the $(x_2, y)$~plane 
for both schemes in \dAu\ collisions. At a fixed value of~$y$, they give 
quite different distributions of the Bjorken-$x$ of the initial gluons. 
The momentum of the final-state gluon considered in the extrinsic scheme 
results in a larger $x_2$-range, 
whereas the intrinsic scheme much heavily favours a tighter band at low~$x_2$ .

\begin{figure}[hbt!]
\includegraphics[height=1.05\columnwidth,angle=90]{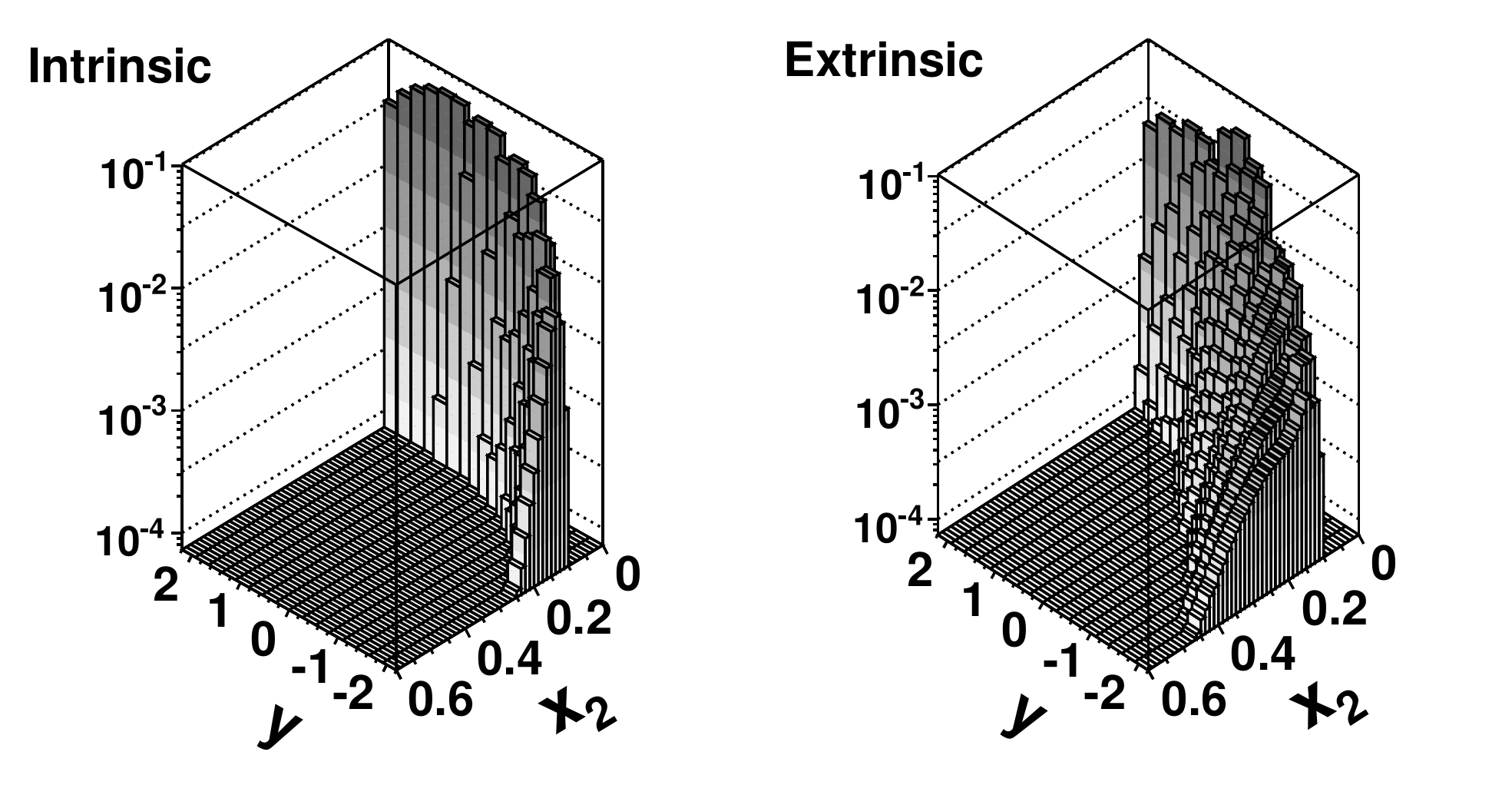}
\caption{Normalised $x_2$ distribution versus rapidity in the intrinsic and extrinsic scenarios for similar Monte-Carlo \jpsi\ statistics in \dAu\ collisions at $\sqrt{s_{NN}}=200\mathrm{~GeV}$.}
\label{fig:intriextriX}
\end{figure}


\section{Results}
\label{sec:results} 


\begin{figure*}[htb!]
\begin{center}
\subfloat[][\RdAu\ versus $y$ in the intrinsic scheme (dashed lines) and extrinsic scheme 
(continuous lines), for several values of the nuclear absorption cross~section~$\sigma_{\mathrm{abs}}$.]{%
\label{fig:RdAu_vs_y-Ncoll-pT-a}
\includegraphics[width=.3\linewidth,angle=90]{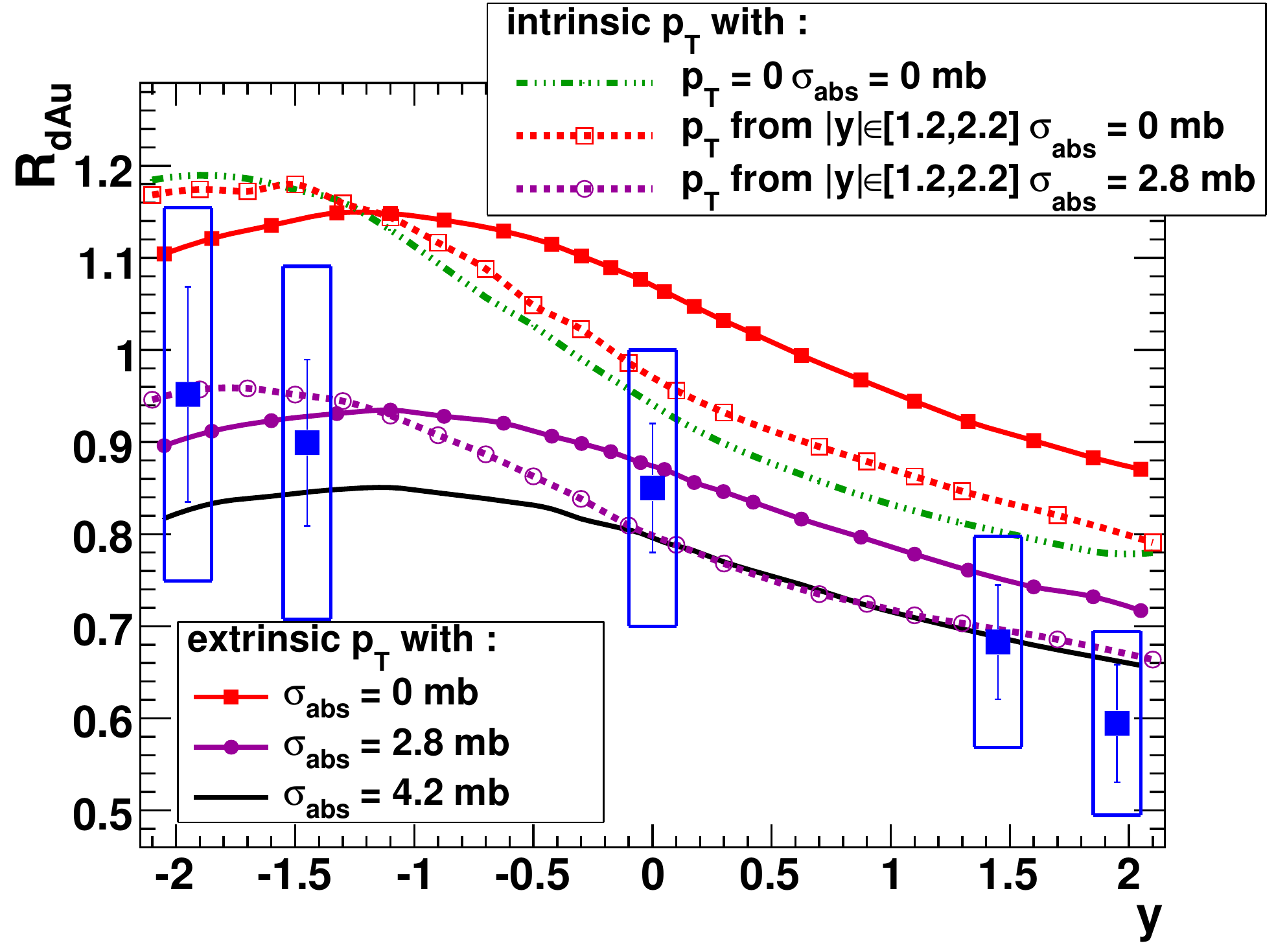}}
\subfloat[][\RdAu\ versus \Ncoll.]{%
\label{fig:RdAu_vs_y-Ncoll-pT-b}
\includegraphics[width=.28\linewidth]{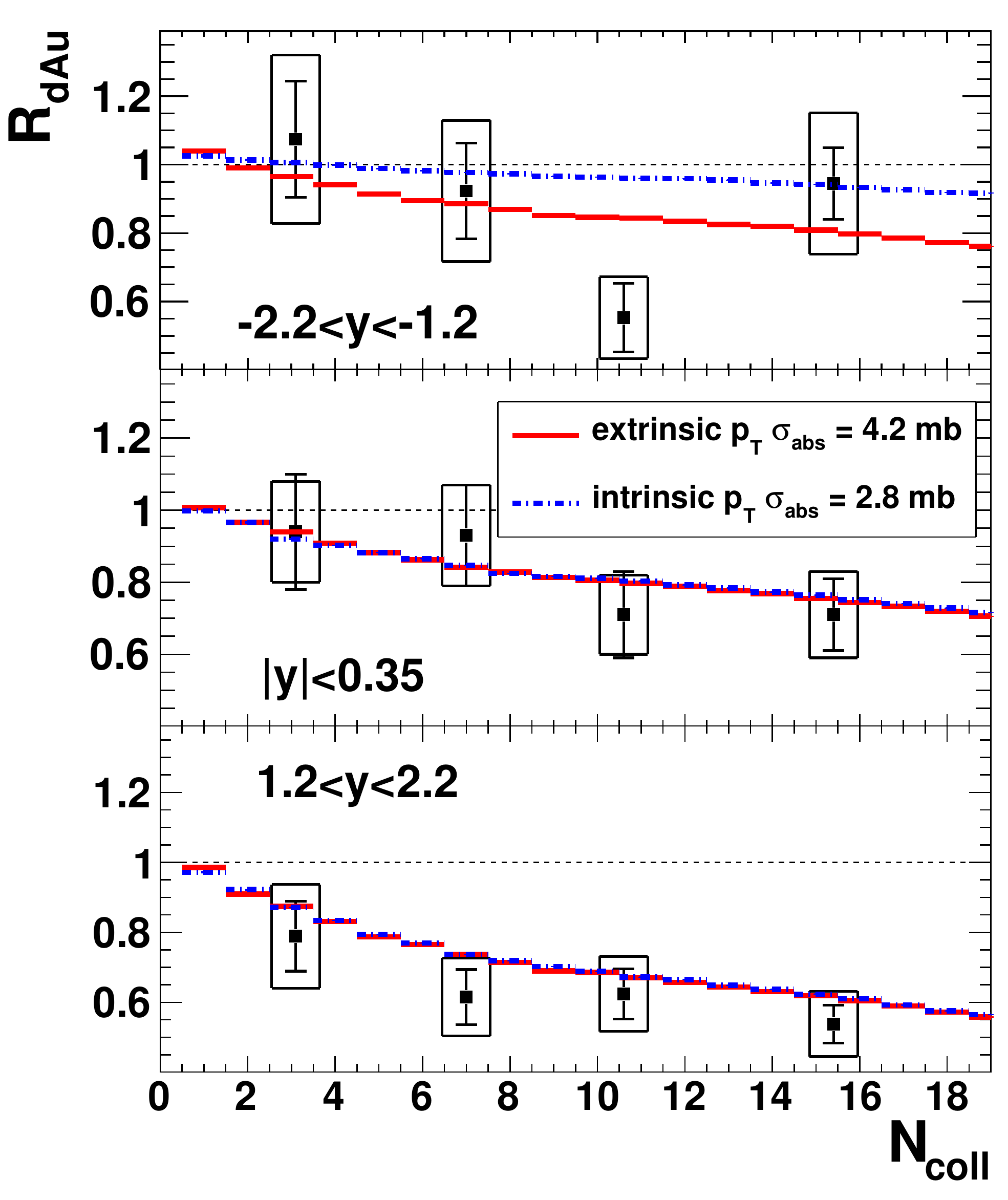}}
\subfloat[][\RdAu\ versus $P_T$.]{%
\label{fig:RdAu_vs_y-Ncoll-pT-c}
\includegraphics[width=.245\linewidth]{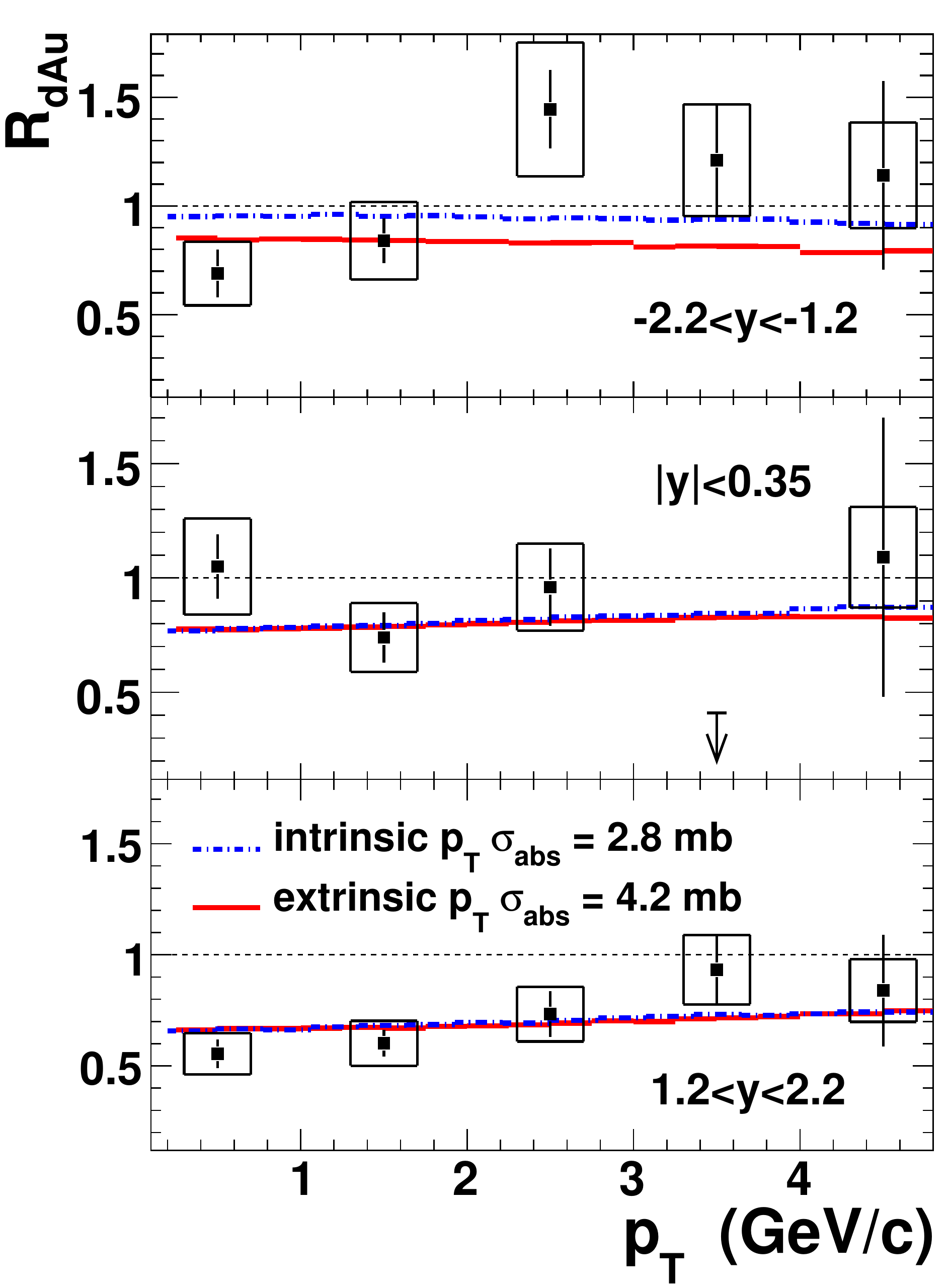}}
\end{center}
\caption{\jpsi\ nuclear modification factor in \dAu\ collisions at $\sqrt{s_{NN}}=200\mathrm{~GeV}$.}
\label{fig:RdAu_vs_y-Ncoll-pT}
\end{figure*}

In the following, we present our results for the  \jpsi\ nuclear modification factor:
\beq
R_{AB}=\frac{dN_{AB}^{J/\psi}}{\langle\Ncoll\rangle dN_{pp}^{J/\psi}}.
\eeq
$dN_{AB}^{J/\psi} (dN_{pp}^{J/\psi})$ is the \jpsi\ yield observed in \AB\ (\pp) collisions 
and $\langle\Ncoll\rangle$ is the average number of nucleon-nucleon collisions occurring 
in one \AB\ collision. In the absence of  nuclear effects, $R_{AB}$ should equal unity.

\subsection{\dAu\ collisions}
\label{subsec:resultsdAu}

\begin{figure*}[htb!]
\vspace{-.5cm}
\begin{center}
\subfloat[][\RAuAu\ in the intrinsic scheme.]{%
\label{fig:RAuAu_vs_Npart_int-ext-a}
\includegraphics[width=.22\linewidth,angle=90]{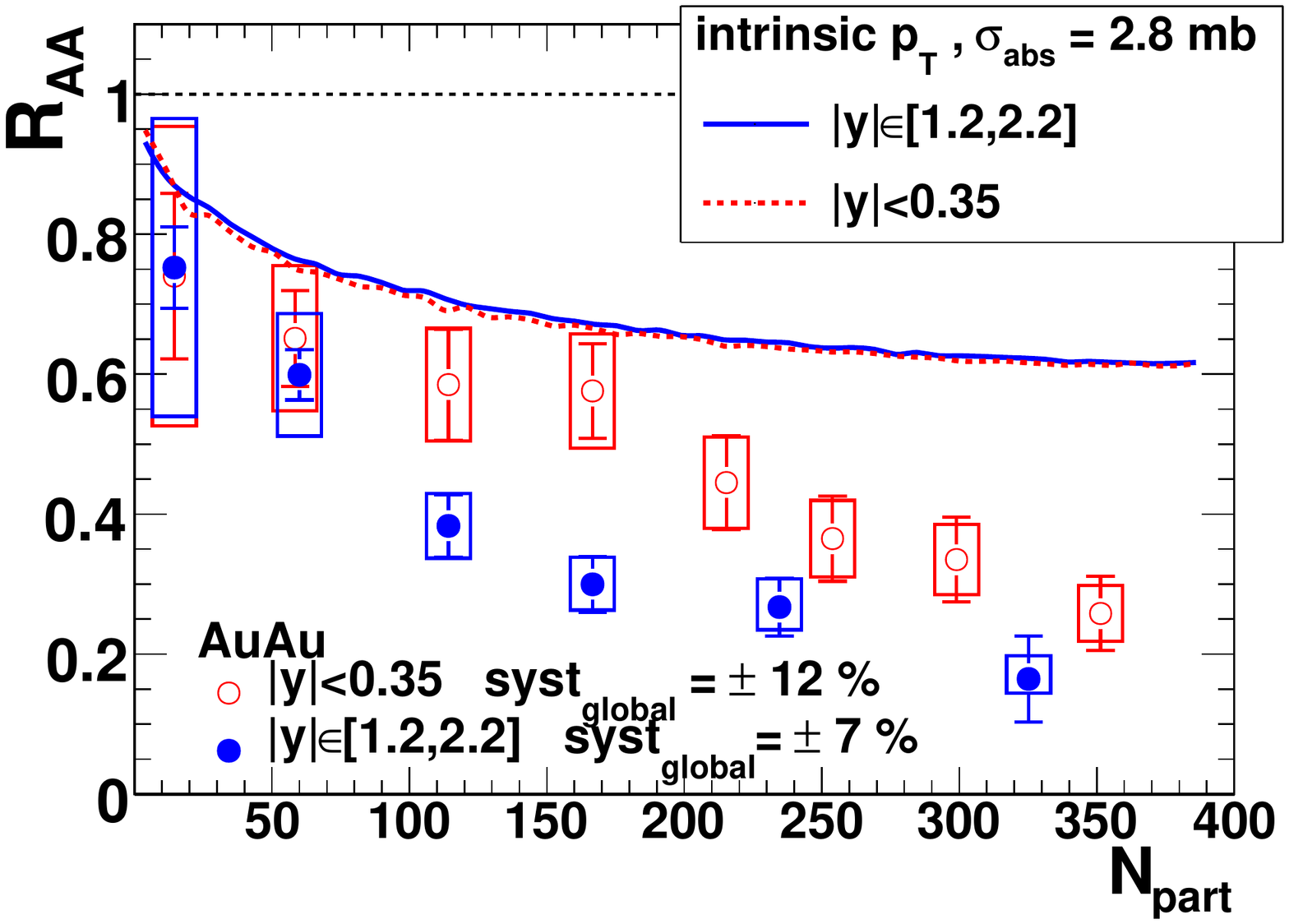}}
\subfloat[][\RAuAu\ in the extrinsic scheme.]{%
\label{fig:RAuAu_vs_Npart_int-ext-b}
\includegraphics[width=.22\linewidth,angle=90]{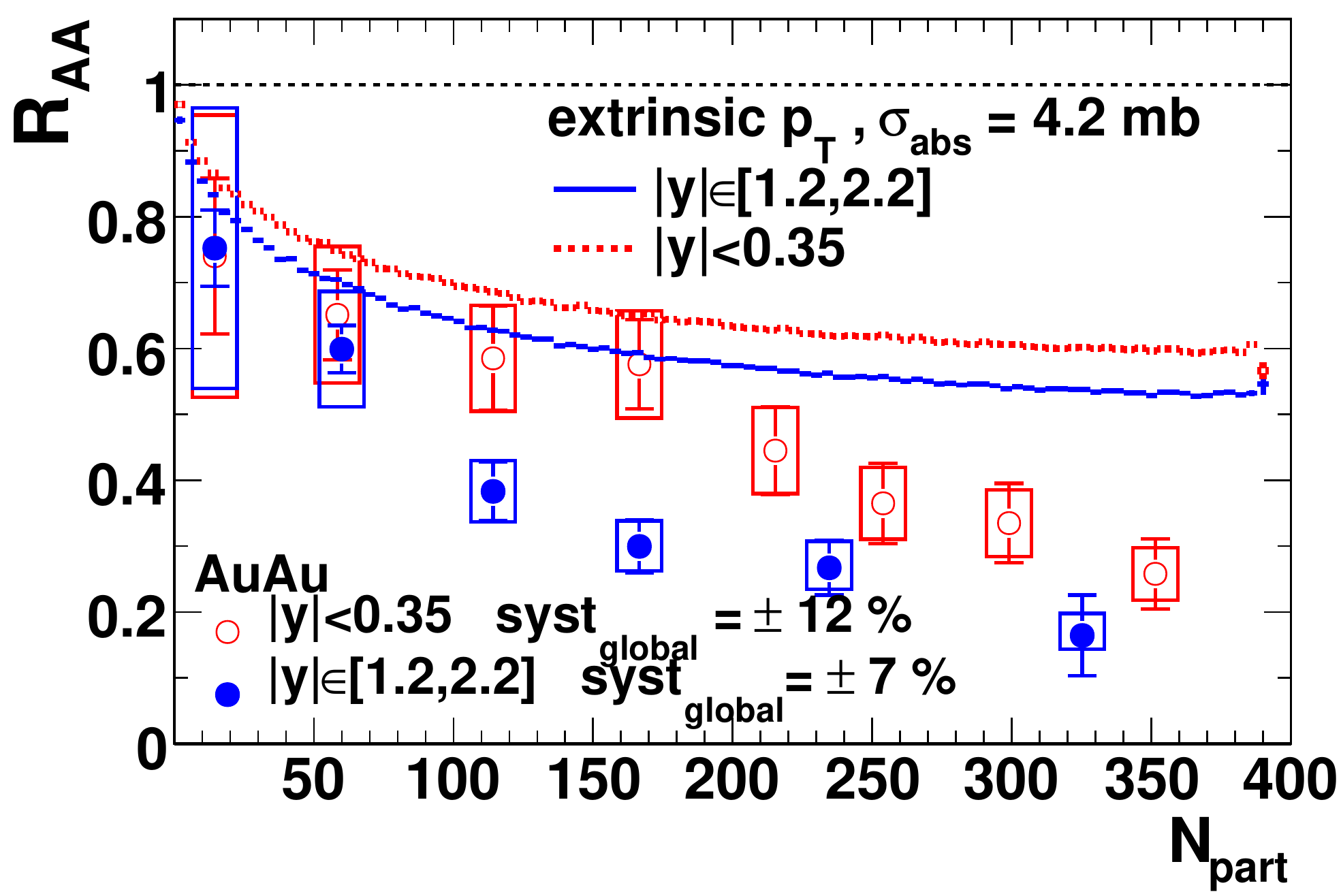}}
\subfloat[][\RCuCu\ in the extrinsic scheme.]{%
\label{fig:RAuAu_vs_Npart_int-ext-c}
\includegraphics[height=.3\linewidth,angle=90]{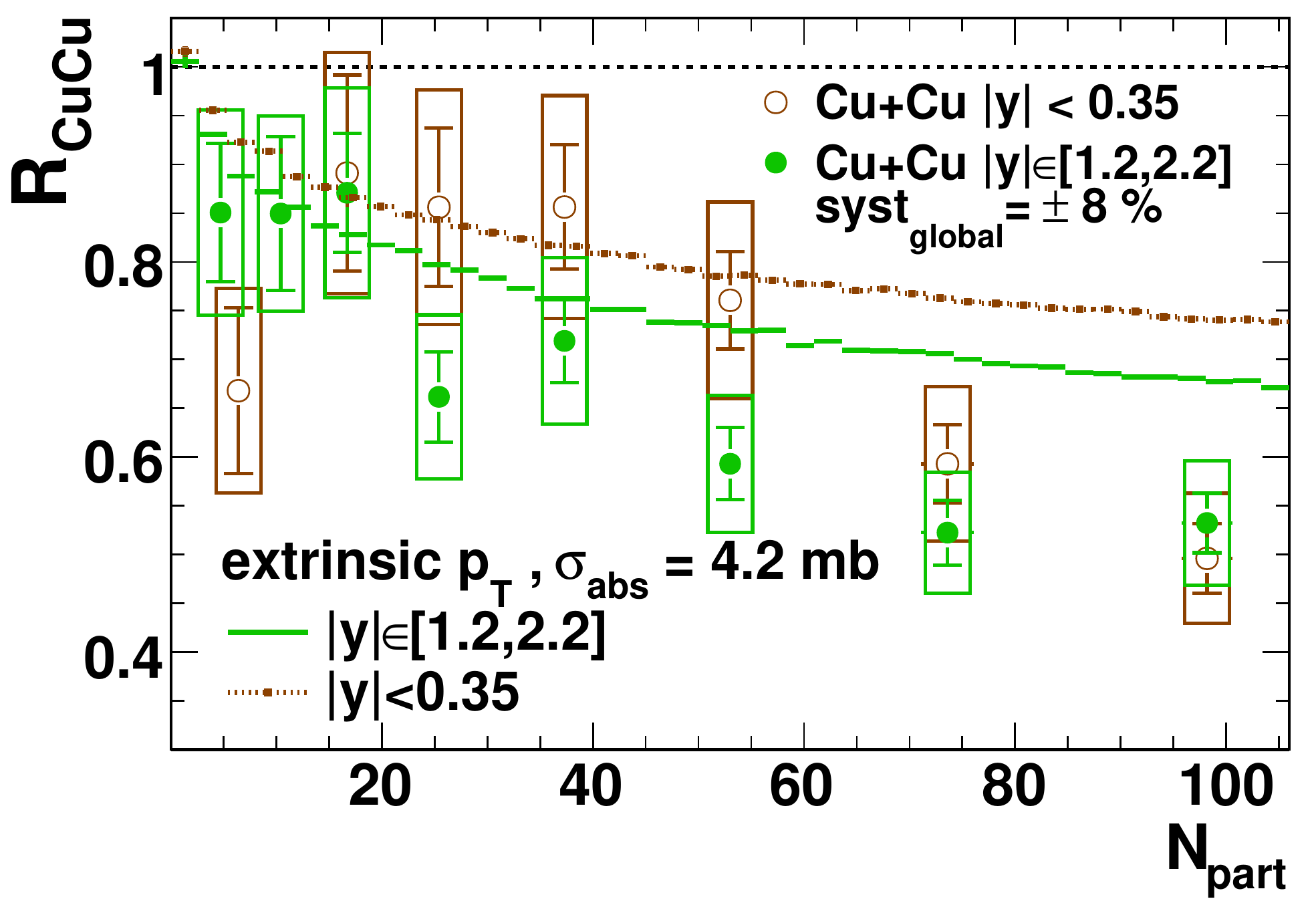}}
\end{center}
\caption{\Npart\ dependence of the \jpsi\ nuclear modification factor in \CuCu\ and \AuAu\ 
collisions at $\sqrtsNN=200\mathrm{~GeV}$.}
\label{fig:RAuAu_vs_Npart_int-ext}
\end{figure*}

PHENIX measurements of \RdAu~\cite{Adare:2007gn} provides with a means to size-up 
the CNM effects at play at RHIC energy. We shall compare the CNM effects obtained in the 
intrinsic and extrinsic schemes to these data.

\cf{fig:RdAu_vs_y-Ncoll-pT-a} shows \RdAu\ versus~$y$. Let us first focus on the curves 
without nuclear absorption. The enhancement due to antishadowing  is shifted to more 
negative~$y$ in the intrinsic scheme\footnote{The dot dashed curve is obtained
for a fixed $\mu_F=M_\psi$ and the dashed one with open squares
for $\mu_F=m_T$ using  the $P_T$ distribution of the $pp$ data
in the forward rapidity region~\cite{OurIntrinsicPaper}. The results are in practice identical if we consider
the distribution in another $y$-region. In the following, all the curves covering the entire rapidity range for the intrinsic
scheme are obtained as this dashed curve.}  compared to the extrinsic scheme. This is easily 
explained. We saw in section~\ref{subsec:see-phase-space} that the typical values 
of $x$ are increased in the extrinsic scheme for given $y$'s. 
A less negative value of $y$ is therefore required to obtain a value 
of $x_2$ producing the maximum amount of antishadowing as seen on~\cf{fig:RdAu_vs_y-Ncoll-pT-a}.

However, as usually done, it is necessary to add the nuclear 
absorption. In the intrinsic scheme, we have used the value of 
$\sigma_{\mathrm{abs}}=2.8\mathrm{~mb}$ from Ref.~\cite{Adare:2007gn}, where it was obtained 
by fitting the data with a shadowing-correction calculation equivalent to neglecting $P_T$ in the 
intrinsic scheme. A higher value of $\sigma_{\mathrm{abs}}$ is required in the extrinsic 
scheme. We used $\sigma_{\mathrm{abs}}=4.2\mathrm{~mb}$ which gives a good 
agreement with PHENIX \dAu\ data\footnote{A complete fitting procedure taking into account
  the experimental errors and their correlations is beyond the scope of this first analysis
of the extrinsic effects. By limiting our analysis to a 
constant value for $\sigma_{\rm abs}$, we also disregard possible formation-time effects.}. 
The resulting curves are also plotted 
on~\cf{fig:RdAu_vs_y-Ncoll-pT-a}. From now on, we will keep these values for the 
value of $\sigma_{\mathrm{abs}}$ in the respective schemes.

Fig.~\ref{fig:RdAu_vs_y-Ncoll-pT-b} shows \RdAu\ as a function of \Ncoll\ in the three rapidity windows. 
Both nuclear shadowing and absorption are included. Both schemes agree
well with the data, especially at mid-$y$

Fig.~\ref{fig:RdAu_vs_y-Ncoll-pT-c} shows \RdAu\ as a function of $P_T$ in the three rapidity windows. 
The intrinsic and extrinsic approaches give similar results. They both give a reasonable agreement 
with the data. 

\subsection{Nucleus-nucleus collisions}
\label{subsec:resultsAA}

We now extend our study to \AA\ collisions, where the \jpsi\ yield may be 
further affected by dense matter effects. In \AuAu\ collisions~\cite{Adare:2006ns}, 
PHENIX has measured an unexpected stronger \jpsi\ suppression at forward-$y$ 
than at mid-$y$. A possible explanation lies in the recombination 
scenarios~\cite{recombniationRefs}. In the following, we shall rather investigate 
if part of this rapidity-dependent suppression may be due to CNM effects. We shall 
also compare our results to PHENIX \CuCu\ data~\cite{Adare:2008sh}. They cover with 
better precision the lower \Npart\ range, where a smaller amount of dense matter 
effects is expected.

\cf{fig:RAuAu_vs_Npart_int-ext-a} shows \RAuAu\ versus \Npart\ in the 
intrinsic scheme, while \cf{fig:RAuAu_vs_Npart_int-ext-b} and~\cf{fig:RAuAu_vs_Npart_int-ext-c}
show \RCuCu\ and \RAuAu\ for the extrinsic scheme. As regards \RCuCu\ and \RAuAu versus $y$, 
they are displayed on \cf{fig:RAuAu_vs_y} for four different centrality bins.

In the intrinsic case, \RAuAu\ is nearly independent of $y$ (\cf{fig:RAuAu_vs_y} up). Indeed, 
for any value of $y$ in the rapidity range $-2 < y < 2$, the suppression at $y>0$ 
in one nucleus is compensated by an enhancement at $y<0$ in the 
other nucleus, thus giving the same result as at mid-rapidity.

On the other hand, in the extrinsic scenario, this cancellation is not as effective
and  both \RCuCu\ and \RAuAu\ show a maximum at $y=0$ (\cf{fig:RAuAu_vs_y}). This is explicit 
in \cf{fig:RAuAu_vs_Npart_int-ext-b} and~\cf{fig:RAuAu_vs_Npart_int-ext-c} where the curves 
for the central and forward rapidity 
ranges are shifted from each other, as the data are.

 \begin{figure}[thb!]
\begin{center}
\includegraphics[width=.9\linewidth]{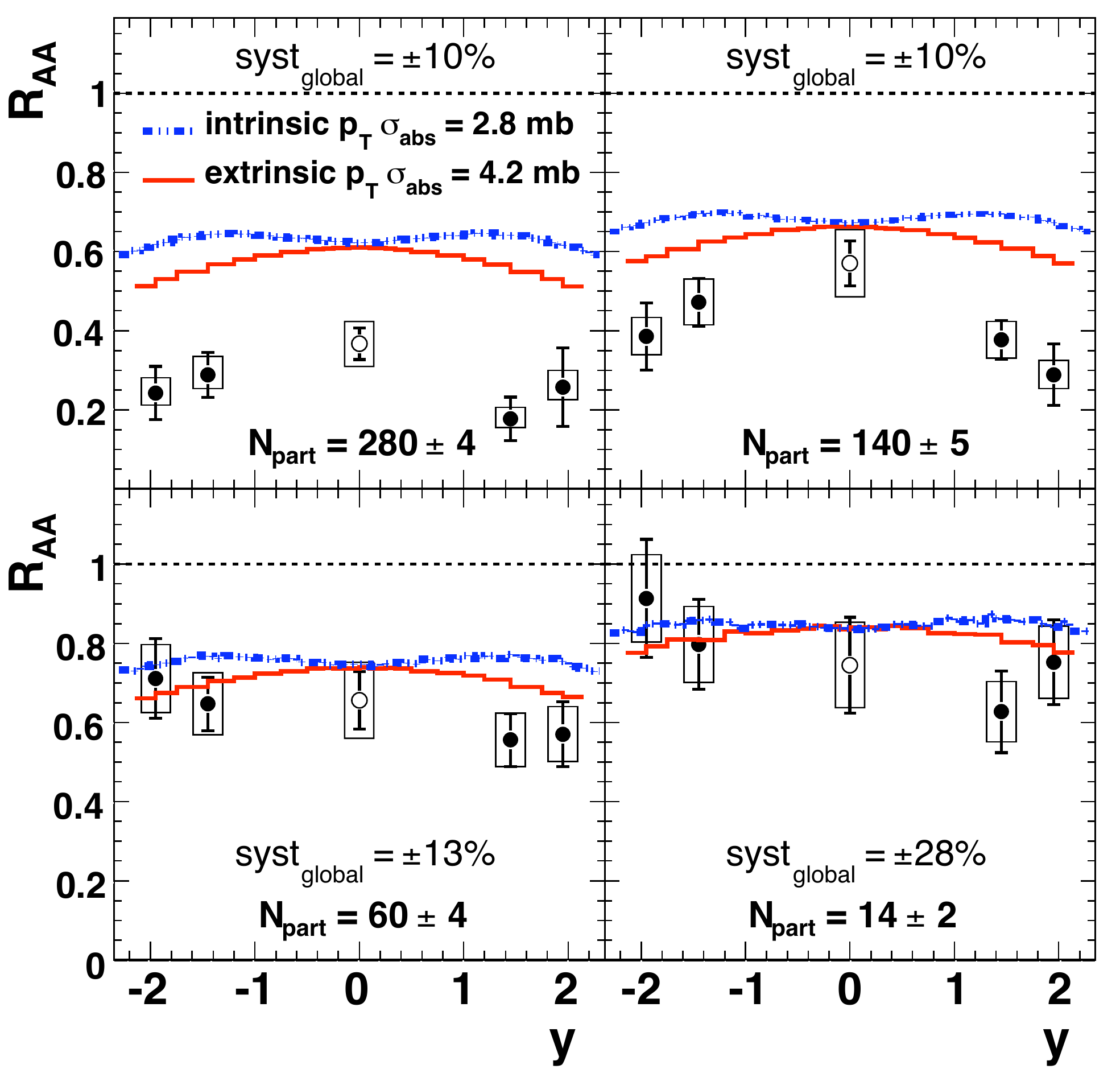}\\
\includegraphics[width=.9\linewidth]{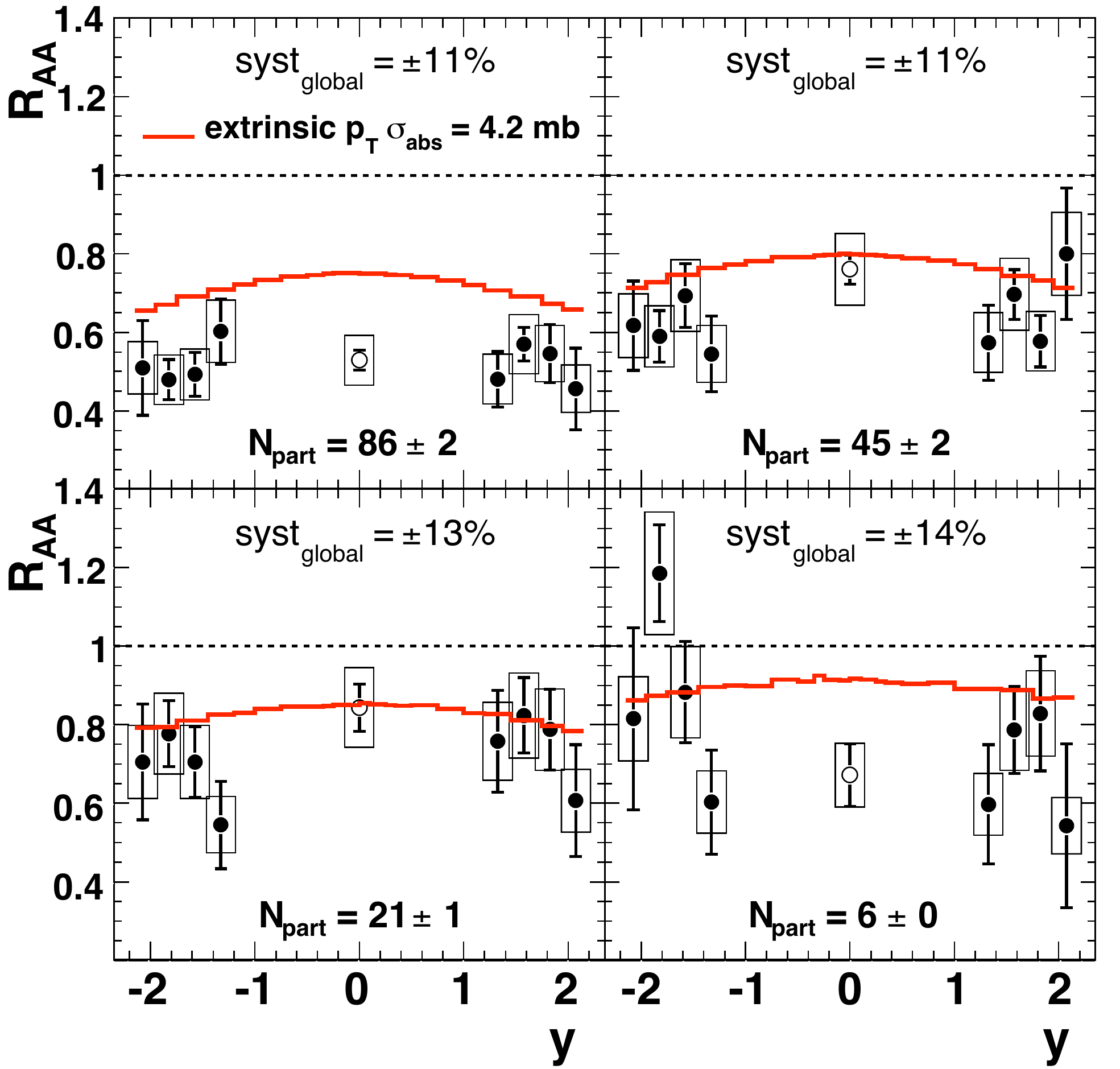}
\end{center}
\caption{\jpsi\ nuclear modification factor versus rapidity in \AuAu\ (top) and \CuCu\ (bottom) collisions.}
\label{fig:RAuAu_vs_y}
\end{figure}


\section{Conclusion and outlook}


We have evaluated Cold Nuclear Matter effects on \jpsi\ production in proton-nucleus 
and nucleus-nucleus collisions at relativistic energies. We have considered the \jpsi\ 
transverse momentum effects within the specific kinematics of the \jpsi\ production at the partonic 
level. We have studied both intrinsic ($2 \to 1$-like process) and extrinsic ($2 \to 2$-like 
process) production schemes. 

We have obtained  different gluon-shadowing-induced effects in \dAu\, depending on 
the considered scheme. We have then included the nuclear absorption cross section needed 
to reproduce PHENIX \dAu\ data~\cite{Adare:2007gn}. In the simplified kinematics 
of a $2\to 1$ process (previously considered in~\cite{OtherShadowingRefs,Vogt:2004dh,OurIntrinsicPaper}), 
we have used $\sigma_{\mathrm{abs}}=2.8\mathrm{~mb}$ according to~\cite{Adare:2007gn}. 
Within the extrinsic scheme, a larger break-up cross section is needed and we used 
$\sigma_{\mathrm{abs}}=4.2\mathrm{~mb}$.

Concerning nucleus-nucleus collisions, we have studied both \CuCu\ and \AuAu\ collisions
 in order to compare our prediction with PHENIX measurements~\cite{Adare:2006ns,Adare:2008sh}. 
Within the extrinsic scheme, we observed a rapidity dependence of \RCuCu\ 
and \RAuAu, in the same direction as the one seen in the data.

In the near future, we plan to extend our investigations to the LHC energies, where we could 
consider the production of $\Upsilon$ in \pA\ and \AA\ collisions at nonzero $P_T$ by 
interfacing partonic matrix elements obtained at NLO~\cite{Campbell:2007ws,Artoisenet:2007xi} 
and  NNLO$^\star$\cite{Artoisenet:2008fc} with our code. We could also broaden the present 
study using other partonic matrix elements for \jpsi\ production, by considering
other parametrisations for the shadowing.  A careful comparison with results from the
Colour-Evaporation Model at NLO~\cite{Bedjidian:2004gd} is planned. 
A better treatment of the interaction between 
the $c \bar c$ pair and the nuclear matter could be also achieved by taking into account 
coherence effects such as the $c$-quark shadowing~\cite{Kopeliovich:2001ee}. Studies of 
the $k_T$~broadening are also envisioned.

In summary, we have demonstrated that the kinematics of the partonic processes responsible 
for the \jpsi\ production is of particular relevance to assess the importance of CNM effects
 both in \pA\ and \AA\ at RHIC energies. Moreover, we argue that a significant part of the 
rapidity dependence of $R_{AA}$ in the central collisions can be accounted by CNM effects only.


\section*{Acknowledgments}


We would like to thank S.~J.~Brodsky, J.~Cugnon, O.~Drapier, R. Granier de Cassagnac, 
P.~Hoyer, B.~Kopeliovich, M. Leitch,
C.~Louren\c co, H.~J.~Pirner, 
C.~A.~Salgado, J.~Stachel and R.~Vogt for stimulating and useful discussions. E.~G.~F. 
thanks Xunta de Galicia (2008/012) and Ministerios de Educacion y 
Ciencia of Spain (FPA2008-03961-E/IN2P3) for financial support. This work is supported in
part by a Francqui fellowship of the Belgian American Educational
Foundation and by the U.S. Department of Energy
under contract number DE-AC02-76SF00515.


\end{document}